\documentclass[twocolumn]{aastex631}

\usepackage[dvipsnames]{xcolor}

\newcommand{\msun}{${\rm M_{\odot}}$}
\usepackage{comment}
\usepackage{amsmath,amssymb}
\usepackage{multirow}
\usepackage{booktabs}
\usepackage{CJKutf8}
\usepackage{mathrsfs}

\begin{document}


\title{A Deep Chandra X-ray Survey of a Luminous Quasar Sample at $z\sim$ 7 }

\author[0000-0002-5768-738X]{Xiangyu Jin}
\affiliation{Steward Observatory, University of Arizona, 933 N Cherry Ave, Tucson, AZ 85721, USA}
\affiliation{Department of Astronomy, University of Michigan, 1085 S. University Ave., Ann Arbor, MI 48109, USA}

\author[0000-0002-7633-431X]{Feige Wang}
\affiliation{Department of Astronomy, University of Michigan, 1085 S. University Ave., Ann Arbor, MI 48109, USA}

\author[0000-0001-5287-4242]{Jinyi Yang}
\affiliation{Department of Astronomy, University of Michigan, 1085 S. University Ave., Ann Arbor, MI 48109, USA}

\author[0000-0003-3310-0131] {Xiaohui Fan}
\affiliation{Steward Observatory, University of Arizona, 933 N Cherry Ave, Tucson, AZ 85721, USA}

\author{Fuyan Bian}
\affiliation{European Southern Observatory, Alonso de Córdova 3107, Casilla 19001, Vitacura, Santiago 19, Chile}

\author[0000-0001-6239-3821]{Jiang-Tao Li}
\affiliation{Purple Mountain Observatory, Chinese Academy of Sciences, 10 Yuanhua Road, Nanjing 210023, People’s Republic of China}

\author[0000-0003-3762-7344]{Weizhe Liu \begin{CJK}{UTF8}{gbsn}(刘伟哲)\end{CJK}}
\affiliation{Steward Observatory, University of Arizona, 933 N Cherry Ave, Tucson, AZ 85721, USA}

\author[0000-0003-4247-0169]{Yichen Liu}
\affiliation{Steward Observatory, University of Arizona, 933 N Cherry Ave, Tucson, AZ 85721, USA}

\author[0000-0002-6221-1829]{Jianwei Lyu}
\affiliation{Steward Observatory, University of Arizona, 933 N Cherry Ave, Tucson, AZ 85721, USA}

\author[0000-0003-4924-5941]{Maria Pudoka}
\affiliation{Steward Observatory, University of Arizona, 933 N Cherry Ave, Tucson, AZ 85721, USA}

\author[0000-0003-0747-1780]{Wei Leong Tee}
\affiliation{Steward Observatory, University of Arizona, 933 N Cherry Ave, Tucson, AZ 85721, USA}

\author[0000-0003-0111-8249]{Yunjing Wu}
\affiliation{Department of Astronomy, Tsinghua University, Beijing 100084, China}

\author[0000-0002-4321-3538] {Haowen Zhang}
\affiliation{Steward Observatory, University of Arizona, 933 N Cherry Ave, Tucson, AZ 85721, USA}

\author[0000-0003-3307-7525]{Yongda Zhu}
\affiliation{Steward Observatory, University of Arizona, 933 N Cherry Ave, Tucson, AZ 85721, USA}



\begin{abstract}

We present new Chandra observations of seven luminous quasars at $z>6.5$. Combined with archival Chandra observations of all other known quasars, they form nearly complete X-ray observations of all currently known $z\sim7$ quasars with $M_{1450}<-26.5$, except for J0313$-$1806 at $z=7.642$ and J0910$-$0414 at $z=6.636$. Together with existing ground-based NIR spectroscopy and ALMA observations, we investigate the correlations between X-ray emission (the X-ray luminosity $L_{\rm X}$ and the optical/UV-to-X-ray spectral slope $\alpha_{\rm OX}$) and various quasar properties (rest-UV luminosity $L_{\rm 2500\;\!\AA}$, bolometric luminosity $L_{\rm bol}$, \ion{C}{4} blueshift, and infrared luminosity $L_{\rm IR}$). We find most $z>6.5$ quasars follow a similar $\alpha_{\rm OX}-L_{\rm 2500\;\!\AA}$ relation as $z\sim1-6$ quasars, but also display a large scatter. 
We find a potential correlation between $\alpha_{\rm OX}$ and the \ion{C}{4} blueshift, suggesting a soft optical/UV-to-X-ray SED shape is frequently associated with fast disk winds. Furthermore, we analyze the X-ray spectrum of 11 quasars at $z>6.5$ with Chandra detection, and find the best-fit photon index $\Gamma$ is $2.41\pm0.27$, which is likely driven by high accretion rates of $z>6.5$ quasars. In addition, we find there are no significant correlations between either $L_{\rm X}$ and $L_{\rm IR}$, nor $L_{\rm bol}$ and $L_{\rm IR}$, suggesting no strong correlations between quasar luminosity and star formation luminosity for the most luminous quasars at $z>6.5$.

\end{abstract}



\section{Introduction} \label{sec:intro}

With extensive high-redshift quasar surveys over twenty years \cite[e.g.,\,][]{Fan2001AJ,Banados2016ApJS,Jiang2016ApJ,Reed2017MNRAS,Wang2017ApJ,Wang2019ApJ,Yang2017AJ,Yang2019AJ,Yang2023ApJS}, more than 200 $z>6$ quasars have been discovered so far \cite[for a review, see][]{Fan2023ARAA}. 
Using ground-based near-infrared (NIR) and JWST spectroscopy \cite[e.g.,][]{Jiang2007AJ,Willot2010AJ,DeRosa2014ApJ,Mazzucchelli2017ApJ,Onoue2019ApJ,Shen2019ApJ,Schindler2020ApJ,Yang2021ApJ,Yang2023ApJb,Loiacono2024AA,Marshall2024arXiv}, a substantial proportion of $z>6$ quasars have been found to host $>10^{9}\;\!$\msun\ supermassive black holes (SMBHs). The existence of billion-solar-mass SMBHs when the universe was less than 1~Gyr years old puts strong constraints on 
the formation and growth of SMBH seeds. To unravel the mystery of the growth of early quasars at $z>6$, the key is to investigate their accretion processes. 

X-ray emission from quasars is a distinctive signature of active black hole accretion. Being highly penetrative, X-ray can reveal active black hole activity even for obscured ($N_{\rm H}>10^{22}\;\!{\rm cm^{-2}}$) quasars \cite[e.g.,][]{HA2018ARAA,Bogdan2024NatAs}.  
The X-ray continuum emission of accreting SMBHs is primarily characterized by a power-law spectrum with a photon index $\Gamma$ and an exponential cutoff at high energy \cite[e.\,g.,\,][]{Garcia2015ApJ}. 
The power-law continuum originates from inverse Compton scattering of UV/optical photons from the accretion disk, by high-energy electrons in the hot corona surrounding the disk \citep{HM1991ApJ}. 
As such, X-ray emission from quasars encodes direct information about black hole accretion. 
%


Many studies have investigated X-ray emission from $z\gtrsim6$ quasars \cite[e.g.,][]{Nanni2017AA,Vito2019AA,Medvedev2020MNRAS,Medvedev2021MNRAS,Wang2021ApJ,Yang2022ApJ,Wolf2023AA,Zappacosta2023AA}, 
including X-ray photometry and spectroscopy, as well as their correlations with quasar properties measured from the rest UV/optical spectrum. $z=0-6$ quasars show a tight correlation between the optical/UV-to-X-ray spectral slope ($\alpha_{\rm OX}$) and the UV luminosity at rest frame 2500~${\rm \AA}$ ($L_{\rm 2500\;\!\AA}$) \cite[e.g.,][]{Just2007ApJ,Green2009ApJ,Lusso2010AA,Nanni2017AA,Timlin2020MNRAS}, such that at a higher $L_{\rm 2500\;\!\AA}$, the disk emission becomes more prominent compared with the corona emission, and the quasar optical/UV-to-X-ray spectral energy distribution (SED) becomes softer. 
With $z\gtrsim6$ quasar samples, no significant redshift evolution has been found in the $\alpha_{\rm OX}-L_{\rm 2500\;\!\AA}$ relation up to $z\sim7$ \citep{Nanni2017AA,Vito2019AA,Wang2021ApJ,Zappacosta2023AA}, suggesting the optical/UV-to-X-ray SED of quasars does not evolve significantly with redshift \citep{Brandt2015ARAA}. 


However, the photon index $\Gamma$ measured from $z>6$ quasar X-ray spectroscopy is $\sim2.2-2.7$, significantly higher than $\Gamma\sim1.8-2.0$ measured among $1<z<6$ quasars \citep{Vito2019AA,Wang2021ApJ,Medvedev2021MNRAS,Yang2022ApJ,Zappacosta2023AA}. 
Some studies argue for a redshift evolution in the accretion properties from $z>6$ bright quasars to $z<6$ quasars \cite[e.g.,][]{Zappacosta2023AA}, but a few other studies suggest that high $\Gamma$ is mainly driven by the high accretion rate of luminous quasars \cite[e.\,g.,\,][]{Vito2019AA,Wang2021ApJ}. Nevertheless, existing X-ray data of quasars at $z>6$ include only a limited number of sources, hindering the characterization of the X-ray properties of the early luminous quasar population. 

There are a few lines of observational evidence indicating that quasars at $z\gtrsim6$ display more frequent and more powerful winds compared with their lower-redshift analogs,  
including a higher fraction of broad-absorption line (BAL) quasars \citep{Bischetti2022Natur}, and a higher average \ion{C}{4} blueshift velocity \citep{Mazzucchelli2017ApJ,Meyer2019MNRAS,Schindler2020ApJ,Yang2021ApJ}. With JWST NIRCam, NIRSpec and ALMA observations, a number of quasars at $z\gtrsim6$ also show the signatures of galactic scale outflows \citep{Marshall2023AA,Yang2023ApJb,Decarli2024AA,Liu2024ApJ,Liu2025arXiv,Lyu2024arXiv,Spilker2025ApJ,Zhu2025arXiv}. 
X-ray emission from low-redshift quasars has been found to be tightly related with existence of powerful winds \cite[e.g.,][]{Richards2011AJ}. Among $z\sim2$ quasars, quasars with a high \ion{C}{4} blueshift velocity usually display a softer $\alpha_{\rm OX}$ \cite[e.g.,][]{Richards2011AJ,Timlin2020MNRAS}. 
Investigations of the relation between X-ray emission and existence of winds have been limited to only a few $z>6$ quasar samples \citep{Wang2021ApJ,Zappacosta2023AA,Tortosa2024AA}, with a tentative trend toward a softer $\alpha_{\rm OX}$ with a higher \ion{C}{4} blueshift. As such, statistical quasar observations are required to investigate the relation between X-ray emission and the existence of quasar winds.  

In this paper, we present new Chandra ACIS-S observations of seven luminous quasars at $z>6.5$. 
Together with existing Chandra observations of other 11 quasars, they form the first statistically complete X-ray observations of all known $z>6.5$ quasars with the absolute rest-frame 1450~${\rm \AA}$ magnitude  $M_{\rm 1450} < -26.5$, except for J0313$-$1806 at $z=7.642$ \citep{Wang2021ApJj0313} and J0910$-$0414 at $z=6.636$ \citep{Wang2019ApJ,Wang2024ApJ}. The Chandra observations of J0910$-$0414 will be reported in separate papers. 
Apart from Chandra observations, all these quasars have been extensively studied with ground-based NIR observations from 6$-$10m telescopes \citep{Yang2021ApJ} and Atacama Large Millimeter Array (ALMA) observations \citep{Wang2024ApJ}, enabling multi-wavelength analysis of quasar and host galaxy properties. Throughout this paper, we adopt a flat $\Lambda$CDM cosmology model with $H_{0}=70~{\rm km\;\!s^{-1}\;\!Mpc^{-1}}$, $\Omega_{\rm m}=0.3$ and $\Omega_{\rm \Lambda}=0.7$. Quoted uncertainty is at the 68\% confidence level and quoted energy is in the observed frame, unless otherwise noted.




\section{Data Reduction} \label{sec:data}
\subsection{Chandra X-ray Observations}
New Chandra ACIS-S observations of seven luminous quasars at $z>6.5$ are obtained in Cycle 22 \cite[Proposal number: 22700552, PI: X. Fan, ][]{Fan2020cxo}. These quasars were first selected as high-redshift quasar candidates through wide area imaging survey, and were later confirmed through spectroscopic observations (\citealt{Wang2018ApJ,Wang2019ApJ}, see also \citealt{Matsuoka2018ApJS}, and \citealt{Yang2019AJ,Yang2020ApJ}). 
The total exposure time of seven quasars is 702.5~ks. For each quasar, the total exposure time is from 50~ks to 120~ks. Each quasar is located at the default aimpoint of the ACIS-S3 chip. 
All observations are in very faint (VFAINT) mode. We summarize all ACIS-S observations 
in Table \ref{tab:XrayObs_Information}.

\begin{figure*}[!h]
    \centering
    \includegraphics[width=1.0\textwidth]{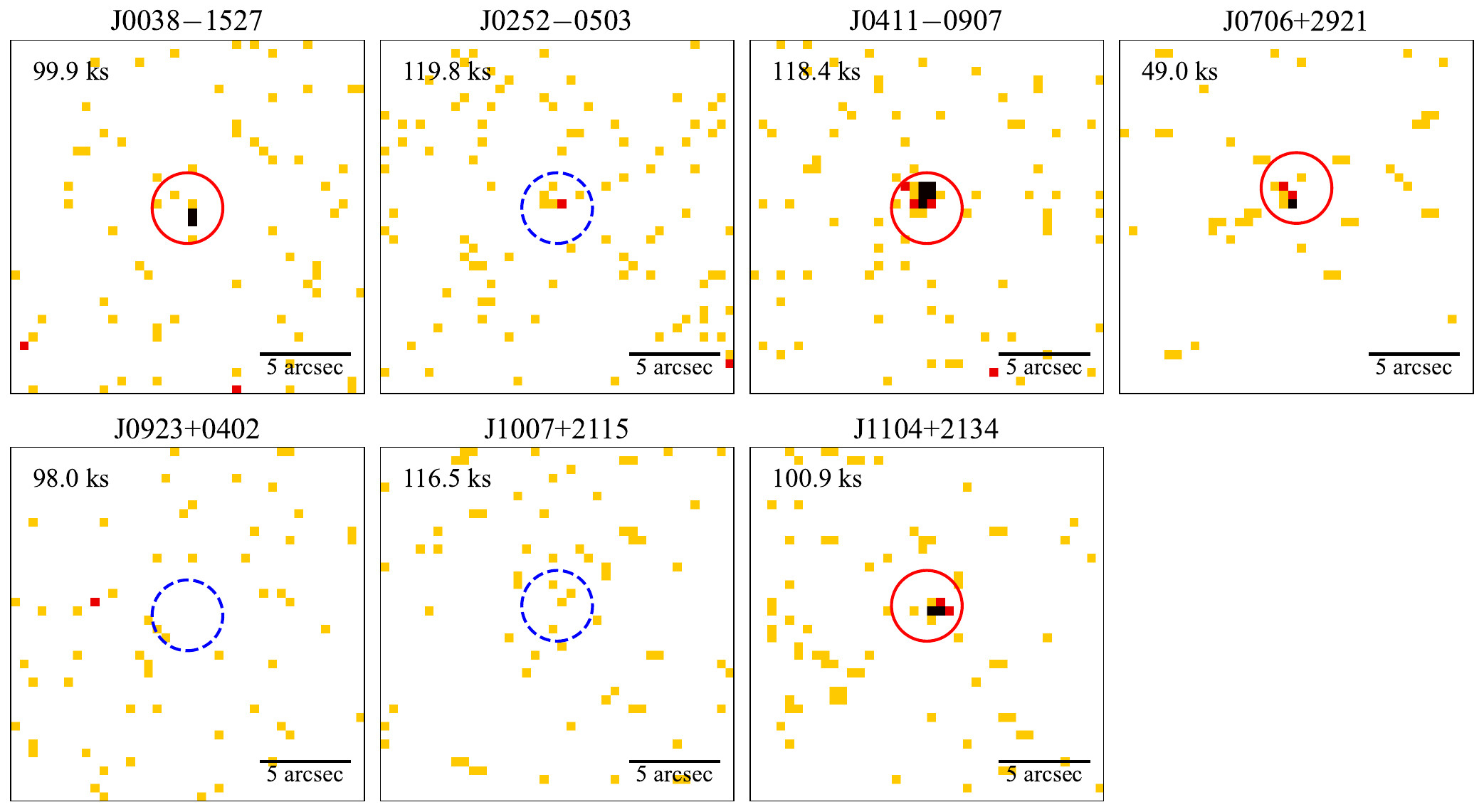}
    \caption{The merged $0.5-7.0$~keV count map of quasars presented in this work. The circular source region centered at the optical position of the quasar is marked by a circle with a radius of $2\arcsec$. 
    For sources showing $\gtrsim2\sigma$ excesses in $0.5-7$~keV net counts within their source regions and binomial no-source probability $P_{\rm B} <0.01$, the circle is red.
    For the undetected source, the circle is in the blue dashed line. The exposure time in kiloseconds (ks) for each stacked count map is indicated in the upper left corner.}
    \label{fig:chandra_cutout}
\end{figure*}

We use \texttt{ciao-4.17} and CALDB version of 4.12.2 to reduce \textit{Chandra} observations. We first reprocess all the observations with \texttt{chandra\_repro} and set \texttt{check\_vf\_pha=yes}. We use \texttt{fine\_astro} command to correct the relative astrometry between multiple Chandra exposures of the same target. \texttt{fine\_astro} command first runs \texttt{wavdetect} on individual exposures to obtain a source list of each exposure. \texttt{fine\_astro} uses the source list of the longest exposure as a reference source list, cross-matches the reference source list with source lists detected in the other exposures, and correct the relative offset between the longest exposure and the other exposure(s). After correcting the astrometry, for each quasar field, we select the optical position of the quasar as a tangent point for stacking. We use \texttt{merge\_obs} with \texttt{psfecf=0.5} and \texttt{psfmerge=expmap} to reproject all existing observations to the tangent point on the original pixel size of $0.492\arcsec$. We then obtain merged count maps, exposure maps and point-spread function (PSF) maps in $0.5-2.0$, $2.0-7.0$, $0.5-7.0$~keV for each quasar field. In these three bands, we use \texttt{wavdetect} to perform source detection on the merged count map on a pixel scale of 1, 2, 4, and 8, and a false source detection rate of $10^{-6}$. 
This step aims to examine whether the target quasar is detected in the merged observation, and to assist in selecting a ``clean" background region free from contamination by any detected X-ray sources, for following analysis. Among seven quasars, J0411$-$0907 
and J1104$+$2134 were detected by \texttt{wavdetect} in all three bands, and J0706$+$2921 was detected by \texttt{wavdetect} in $0.5-2.0$~keV and $0.5-7.0$~keV. Although J0038$-$1527, J0252$-$0503, and J1007$+$2115 were not detected by \texttt{wavdetect} in $0.5-2$ keV, $2-7$ keV or $0.5-7.0$ keV, subsequent \texttt{srcflux} measurements show their nets counts are significant at $\gtrsim2\sigma$ levels in $0.5-2$~keV and $0.5-7.0$~keV, given the source and background regions (see Section \ref{sec:analysis} for more details). Figure \ref{fig:chandra_cutout} shows the cutout images of the 0.5$-$7~keV merged count map, centered at the optical locations of seven quasars. 

\begin{deluxetable*}{ccccc}\centering
\tabletypesize{\scriptsize}
\tablecaption{Summary of Chandra ACIS-S Observations}
\tablewidth{2\columnwidth}
\tablehead{\colhead{Quasar Name} &
\colhead{Redshift $z_{\rm [C\;\!II]}$} & \colhead{Chandra ObsID} & \colhead{Exposure Time (ks)} & \colhead{Chandra Obs. Date}}
\colnumbers
\startdata
\multirow{4}{*}{J0038$-$1527} & \multirow{4}{*}{$7.0340\pm0.0003$} & 23831 & 25.73 & 2020-10-01  \\
             &      & 24275 & 24.73 & 2020-10-03  \\
             &      & 24276 & 22.75 & 2020-10-02  \\
             &      & 24669 & 26.71 & 2020-10-02  \\ \hline
\multirow{7}{*}{J0252$-$0503} & \multirow{7}{*}{$7.0006\pm0.0009$} & 23832 & 15.86 & 2020-10-15 \\
             &      & 24472 & 29.67 & 2020-10-12 \\
             &      & 24473 & 13.88 & 2020-10-15 \\
             &      & 24474 & 19.79 & 2020-10-23 \\
             &      & 24836 & 14.87 & 2020-10-14 \\
             &      & 24837 & 15.86 & 2020-10-17 \\
             &      & 24844 & 9.83  & 2020-11-02 \\ \hline
\multirow{4}{*}{J0411$-$0907} & \multirow{4}{*}{$6.8260\pm0.0007$} & 23833 & 28.39	& 2021-10-15 \\
             &      & 24260 & 29.67 & 2021-11-12 \\
             &      & 24261	& 31.65 & 2021-10-07 \\
             &      & 24262 & 28.69 & 2021-10-26 \\ \hline
\multirow{2}{*}{J0706$+$2921} & \multirow{2}{*}{$6.6037\pm0.0003$} & 23834 & 26.23 & 2020-12-31 \\
             &      & 24917	& 22.77	& 2021-01-08 \\ \hline
\multirow{4}{*}{J0923$+$0402} & \multirow{4}{*}{$6.6330\pm0.0003$} & 23835 & 39.46 & 2022-05-16 \\
             &      & 24339 & 24.74 & 2021-02-10 \\
             &      & 24340	& 7.96 & 2021-02-14 \\
             &      & 24956	& 25.82	& 2021-02-14 \\ \hline
\multirow{6}{*}{J1007$+$2115} & \multirow{6}{*}{$7.5153\pm0.0005$} & 23836	& 20.79	& 2021-02-13 \\
             &      & 24499 & 20.79 & 2021-01-22 \\
             &      & 24500 & 28.19	& 2021-02-16 \\
             &      & 24501	& 24.74	& 2021-02-11 \\
             &      & 24932 & 8.20 & 2021-01-22 \\
             &      & 24955	& 13.78 & 2021-02-14 \\ \hline
\multirow{4}{*}{J1104$+$2134} & \multirow{4}{*}{$6.7662\pm0.0009$} & 23837 & 41.51 & 2022-07-06 \\
             &      & 24477 & 17.83 & 2021-02-27 \\
             &      & 24478 & 29.67	& 2022-05-17 \\
             &      & 24970 & 11.91 & 2021-02-26 \\
\enddata
\tablecomments{(1) Quasar name; (2) Redshift $z_{\rm [C\;\!II]}$ is measured based on the [\ion{C}{2}] 158 \textmu m detection from ALMA observations \citep{Yang2021ApJ,Wang2024ApJ}. (3)--(5) Observation ID (ObsID), exposure time in kiloseconds (ks), and observation date of Chandra observations.}
\end{deluxetable*}\label{tab:XrayObs_Information}

\subsection{Ground Based Near-Infrared Observations}

All quasars included in the \textit{Chandra} program have been extensively studied using ground-based NIR observations. We summarize quasar properties measured from existing ground-based NIR observations in Table \ref{tab:NIRObs_Information}, including black hole mass $M_{\rm BH}$, the absolute rest-frame 1450~${\rm \AA}$ magnitude $M_{\rm 1450}$, bolometric luminosity $L_{\rm bol}$, Eddington ratio $\lambda_{\rm Edd}$, \ion{Mg}{2} redshift and \ion{C}{4} redshift. The NIR spectrum was fitted with a power-law continuum, a \ion{Fe}{2} template, a Balmer continuum, and emission lines including \ion{Fe}{2}, \ion{C}{4}, and \ion{Si}{4} and \ion{C}{3}]. The details of NIR spectral fitting and quasar properties measurements can be found in \citet{Wang2018ApJ,Wang2020ApJ} and \citet{Yang2021ApJ}. 
    
We briefly summarize the spectral fitting procedures below. The NIR spectrum is fitted with the pseudo continuum and emission lines. The pseudo continuum consists of a power-law continuum, a \ion{Fe}{2} template, and a Balmer continuum. After subtracting the best-fit continuum, Gaussian components for \ion{C}{4} and \ion{Mg}{2} (\ion{Si}{4}, and \ion{C}{3} if visible) are fitted for the continuum-subtracted spectrum. The absolute magnitude at rest-frame 1450~${\rm \AA}$ $M_{1450}$ is calculated based on the best-fit power-law continuum at rest-frame 1450~${\rm \AA}$, and the bolometric luminosity $L_{\rm bol}$ is calculated from the rest-frame 3000~${\rm \AA}$ luminosity by assuming a bolometric correction of 5.15 \citep{Richards2006ApJS,Shen2011ApJS}. The black hole mass $M_{\rm BH}$ is calculated using single-epoch method through the \ion{Mg}{2} line: $M_{\rm BH}=10^{6.86}\times[\frac{\rm FWHM (Mg\;\!II)}{1000~{\rm km\;\!s^{-1}}}]^{2}\;\![\frac{\lambda L_{\lambda}(3000{\rm \AA})}{10^{44}~{\rm erg\;\!s^{-1}}}]^{0.5}~M_{\odot}$\citep{VO2009ApJ}. The Eddington ratio $\lambda_{\rm Edd}$ is calculated based on the bolometric luminosity and the black hole mass, using $\lambda_{\rm Edd}=L_{\rm bol}/(1.26\times10^{38}\;\!({\rm M_{BH}/M_{\odot}})\;\!{\rm erg\;\!s^{-1}})$.

\begin{deluxetable*}{ccccccccc}\centering
\tabletypesize{\scriptsize}
\tablecaption{Quasar Properties Measured from NIR Spectroscopy}
\tablewidth{2\columnwidth}
\tablehead{\colhead{Quasar Name} & \colhead{$z_{\rm C IV}$} & \colhead{$z_{\rm Mg II}$} &
\colhead{${\rm M_{BH}}$ ($10^{9}\;\!{\rm M_{\odot}}$)} & \colhead{${\rm M_{1450}}$} & \colhead{Spectral Slope $\alpha_{\lambda}$} & \colhead{$L_{\rm bol}\;\!(10^{47}\;\!{\rm erg\;\!s^{-1}})$} & $\lambda_{\rm Edd}$ &\colhead{Reference}}
\colnumbers
\startdata
J0038$-$1527$^{B}$ & $6.929\pm0.003$ & $6.999\pm0.001$ & $1.36\pm0.05$ & $-27.13$ & $-1.46\pm0.02$ & $2.37\pm0.09$ & $1.4\pm0.1$  & \citet{Wang2018ApJ,Yang2021ApJ} \\ \hline
J0252$-$0503 & $6.867\pm0.005$ & $6.99\pm0.02$ & $1.39\pm0.16$ & $-26.63$ & $-1.67\pm0.04$  & $1.3\pm0.1 $ & $0.7\pm0.1$  &\citet{Wang2020ApJ,Yang2021ApJ} \\ \hline
J0411$-$0907 & $6.790\pm0.005$ & $6.827\pm0.006$ & $0.95\pm0.09$ & $-26.58$ & $-1.31\pm0.03$ & $1.59\pm0.10$ & $1.3\pm0.2$ & \citet{Yang2021ApJ} \\ \hline
J0706$+$2921$^{B}$ & $6.54\pm0.02$ & $6.5925\pm0.0004$ & $2.11\pm0.16$ & $-27.44$ & $-1.35\pm0.03$ & $3.39\pm0.15$ & $1.3\pm0.1$ & \citet{Yang2021ApJ}  \\ \hline
J0923$+$0402$^{B}$ & $6.59\pm0.02$ & $6.612\pm0.002$ & $1.77\pm0.02$ & $-26.68$ & $-1.00\pm0.07$ & $2.17\pm0.03$ & $1.0\pm0.1$ & \citet{Yang2021ApJ} \\ \hline
J1007$+$2115 & $7.39\pm0.04$ & $7.48\pm0.01$ & $1.43\pm0.22$ & $-26.73$ & $-1.14\pm0.01$ & $2.04\pm0.13$ & $1.1\pm0.2$ & \citet{Yang2021ApJ} \\ \hline
J1104$+$2134 & $6.739\pm0.002$ & $6.766\pm0.005$ & $1.69\pm0.15$ & $-26.63$ & $-1.44\pm0.04$ & $1.51\pm0.09$ & $0.7\pm0.1$ & \citet{Yang2021ApJ} \\
\enddata
\tablecomments{(1) Quasar name; (2)-(3) Redshift of \ion{C}{4} ($z_{\rm CIV}$) and \ion{Mg}{2} ($z_{\rm MgII}$) emission lines; (4) Black hole mass ($M_{\rm BH}$); (5) The absolute rest-frame 1450~${\rm \AA}$ magnitude; (6) Spectral slope $\alpha_{\lambda}$ of the power-law continuum; (7) Bolometric luminosity $L_{\rm bol}$; (8) Eddington ratio $\lambda_{\rm Edd}$; (9) Reference for quasar properties.}
\tablenotetext{B}{Broad absorption line (BAL) quasars identified in \citet{Yang2021ApJ}}
\end{deluxetable*}\label{tab:NIRObs_Information}

\section{Measurements} \label{sec:analysis}

For each quasar, we use \texttt{srcflux} to measure net count rates in $0.5-2.0$~keV, $2.0-7.0$~keV, and $0.5-7.0$~keV, and calculate the unabsorbed $0.5-7.0$~keV flux. In \texttt{srcflux}, a power-law model with galactic absorption is assumed to calculate the unabsorbed $0.5-7.0$~keV flux based on the net count rate. We adopt a photon index $\Gamma=2.3$, measured from a joint spectral fit of six $z>6.5$ quasars \citep{Wang2021ApJ}.  
When running \texttt{srcflux}, we adopt a circular source region with a radius of $2\arcsec$, centered at the optical position of the quasar, and an annular background region with an inner radius of $5\arcsec$ and an outer radius of $10\arcsec$, also centered at the quasar optical position. The annular background region does not include detected X-ray sources from \texttt{wavdetect}.
In $0.5-7.0$~keV, we calculate the binomial probability $P_{\rm B}$ associated with non-detection to investigate the detection significance of each source \citep{Weisskopf2007ApJ}: $P_{\rm B} (X\geq S) =\Sigma^{N}_{X=S} \frac{N!}{X!(N-X)!}p^{X}(1-p)^{N-X}$, where $S$ is the counts in the source region, $N$ is the total counts in the source and background regions, and $p=1/(1+bscale)$. $bscale$ is the ratio of the size of background region to the size of the source region. The net counts in $0.5-2.0$~keV, $2.0-7.0$~keV, and $0.5-7.0$~keV, non-detection binomial probability $P_{\rm B}$ in $0.5-7.0$~keV, and unabsorbed $0.5-7.0$~keV flux are reported in Table \ref{tab:Xray_photometry}. For sources with net counts with significance at $<2\sigma$ levels, we report the $2\sigma$ upper limits (95\% confidence intervals). Although the net counts of J0252$-$0503 and J1007$+$2115 are above $2\sigma$ levels in the source region, $P_{\rm B}$ of J0252$-$0503 and J1007$+$2115 is $\gtrsim0.01$, indicating these two sources are marginally detected in $0.5-7.0~$keV. We therefore report the $2\sigma$ upper limits for J0252$-$0503 and J1007$+$2115 as well. 

For each quasar, we use the unabsorbed $0.5-7.0$~keV flux to calculate the rest 2--10~keV luminosity $L_{\rm 2-10~keV}$ and the rest 2~keV monochromatic luminosity $\nu L_{\rm 2~keV}$, 
assuming a photon index of 2.3. 
For undetected sources in 0.5--7.0~keV, we use the $2\sigma$ upper limit of the unabsorbed 0.5--7.0~keV flux to calculate the $2\sigma$ upper limit of $L_{\rm 2-10~keV}$ and $\nu L_{\rm 2~keV}$. We then calculate the optical/UV-to-X-ray spectral slope $\alpha_{\rm OX}$ using $\alpha_{\rm OX}=\frac{{\rm log} (f_{2\;\! \rm keV}/f_{2500 \rm \AA})}{{\rm log} (\nu_{2\;\! \rm keV}/\nu_{2500 \rm \AA})}$ \citep{Tananbaum1979ApJ}, where $f_{2\;\!keV}$ and $f_{2500 \rm \AA}$ are the flux densities at the rest-frame 2~keV and 2500~${\rm \AA}$. $f_{2500 \rm \AA}$ is calculated using $M_{1450}$ and $\alpha_{\rm \lambda}$ measured from ground-based NIR spectroscopy (see Table \ref{tab:NIRObs_Information}). We report $L_{\rm 2-10~keV}$, $\nu L_{\rm 2~keV}$ and $\alpha_{\rm OX}$ of each quasar in Table \ref{tab:Xray_photometry}. 

For quasars with net counts exceeding $2\sigma$ levels in $0.5-7$~keV, we use \texttt{specextract} to extract their X-ray spectrum in individual exposures, adopting the same source and background regions used for \texttt{srcflux}. We obtain the source spectrum, the source auxiliary response file (ARF), the source redistribution matrix file (RMF), the background spectrum, the background ARF, and the background RMF of each individual source through \texttt{specextract}. 

\begin{deluxetable*}{cccccccccc}\centering
\tabletypesize{\scriptsize}
\tablecaption{Quasar X-ray Photometric Properties}
\tablewidth{2\columnwidth}
\tablehead{\colhead{\multirow{2}{*}{Quasar Name}} & \multicolumn{3}{c}{Net Counts}  &  \colhead{\multirow{2}{*}{$P_{\rm B}$}} &
\colhead{\multirow{2}{*}{0.5--7.0~keV Flux}} & \colhead{\multirow{2}{*}{Rest $L_{\rm 2-10~keV}$}} & \colhead{\multirow{2}{*}{Rest $\nu L_{\rm 2~keV}$}} &  \colhead{\multirow{2}{*}{ $\alpha_{\rm OX}$}} & \colhead{\multirow{2}{*}{$k_{\rm bol, X}$}} \\ \cmidrule(lr){2-4} 
& \colhead{0.5--7.0~keV} & \colhead{0.5--2.0~keV} & \colhead{2.0--7.0~keV} &  &  & & & & }
\colnumbers
\startdata
J0038$-$1527 & $7.9_{-2.9}^{+3.5}$ & $5.3_{-2.1}^{+2.8}$ & $<7.8$ & $1.2\times10^{-4}$ & $11.2_{-4.1}^{+4.9}$ & $5.6_{-2.0}^{+2.4}$ & $2.2_{-0.8}^{+0.9}$ & $-1.79_{-0.08}^{+0.06}$ & $426^{+155}_{-187}$\\
J0252$-$0503 & $<11.7$ & $<9.1$ & $<6.3$ & $1.8\times10^{-2}$ & $<14.1$ & $<6.9$ & $<2.7$ & $<-1.65$ & $>188$\\
J0411$-$0907 & $29.5_{-5.3}^{+6.0}$ & $19.8_{-4.3}^{+5.0}$ & $9.6_{-3.0}^{+3.6}$ & $3.3\times10^{-21}$ & $39.0_{-7.1}^{+7.9}$ & $17.7_{-3.2}^{+3.6}$ & $6.9_{-1.3}^{+1.4}$ & $-1.53_{-0.03}^{+0.03}$ & $89^{+17}_{-19}$ \\
J0706+2921 & $9.3_{-2.8}^{+3.5}$ & $6.7_{-2.3}^{+3.0}$ & $2.5_{-1.4}^{+2.1}$ & $6.5\times10^{-8}$ & $29.3_{-9.0}^{+11.2}$ & $12.5_{-3.8}^{+4.8}$ & $4.9_{-1.5}^{+1.9}$ & $-1.71_{-0.06}^{+0.05}$ & $271^{+84}_{-104}$\\
J0923+0402 & $<3.0$ & $<3.0$ & $<3.0$ & 1.0 & $<4.7$ & $<2.0$ & $<0.8$ & $<-1.93$ & $>1084$\\
J1007+2115 & $<10.5$ & $<8.6$ & $<5.5$ & $9.8\times10^{-3}$ & $<13.2$ & $<7.6$ & $<3.0$ & $<-1.71$ & $>267$ \\
J1104+2134 & $15.0_{-3.8}^{+4.4}$ & $7.3_{-2.5}^{+3.2}$ & $7.6_{-2.7}^{+3.4}$ & $6.7\times10^{-10}$ & $22.8_{-5.8}^{+6.8}$ & $10.1_{-2.6}^{+3.0}$ & $3.9_{-1.0}^{+1.2}$ & $-1.62_{-0.05}^{+0.04}$ & $149^{+39}_{-45}$\\
\enddata
\tablecomments{(1) Quasar name; (2)$-$(4) Net counts in observed $0.5-7.0$~keV, $0.5-2.0$~keV, $2.0-7.0$~keV; (5) Binomial probability associated with non-detection in $0.5-7.0$~keV; (6) Unabsorbed flux in the observed $0.5-7.0$~keV, in units of $10^{-16}\;\!{\rm erg\;\!s^{-1}\;\!cm^{-2}}$, assuming a galactic absorption and a power-law index of $2.3$ \citep{Wang2021ApJ}; (7) Luminosity at rest-frame $2-10$~keV, in units of $10^{44}\;\!{\rm erg\;\!s^{-1}}$; (8) Monochromatic luminosity at the rest-frame 2~keV, in units of $10^{44}\;\!{\rm erg\;\!s^{-1}}$; (9) Optical/UV-to-Xray spectral slope $\alpha_{\rm OX}$; (10) X-ray bolometric correction $k_{\rm bol, X}=L_{\rm bol}/L_{\rm 2-10~keV}$.}
\end{deluxetable*}\label{tab:Xray_photometry}

\section{Results and Discussion} \label{sec:discussion}

In this section, we study the X-ray photometric and spectroscopic properties of quasars at $z>6.5$, and investigate their relations with quasar and host galaxy properties measured from NIR spectroscopy and ALMA observations. 

\subsection{X-ray Luminosity versus Bolometric Luminosity}
\begin{figure}
    \centering
    \includegraphics[width=0.5\textwidth]{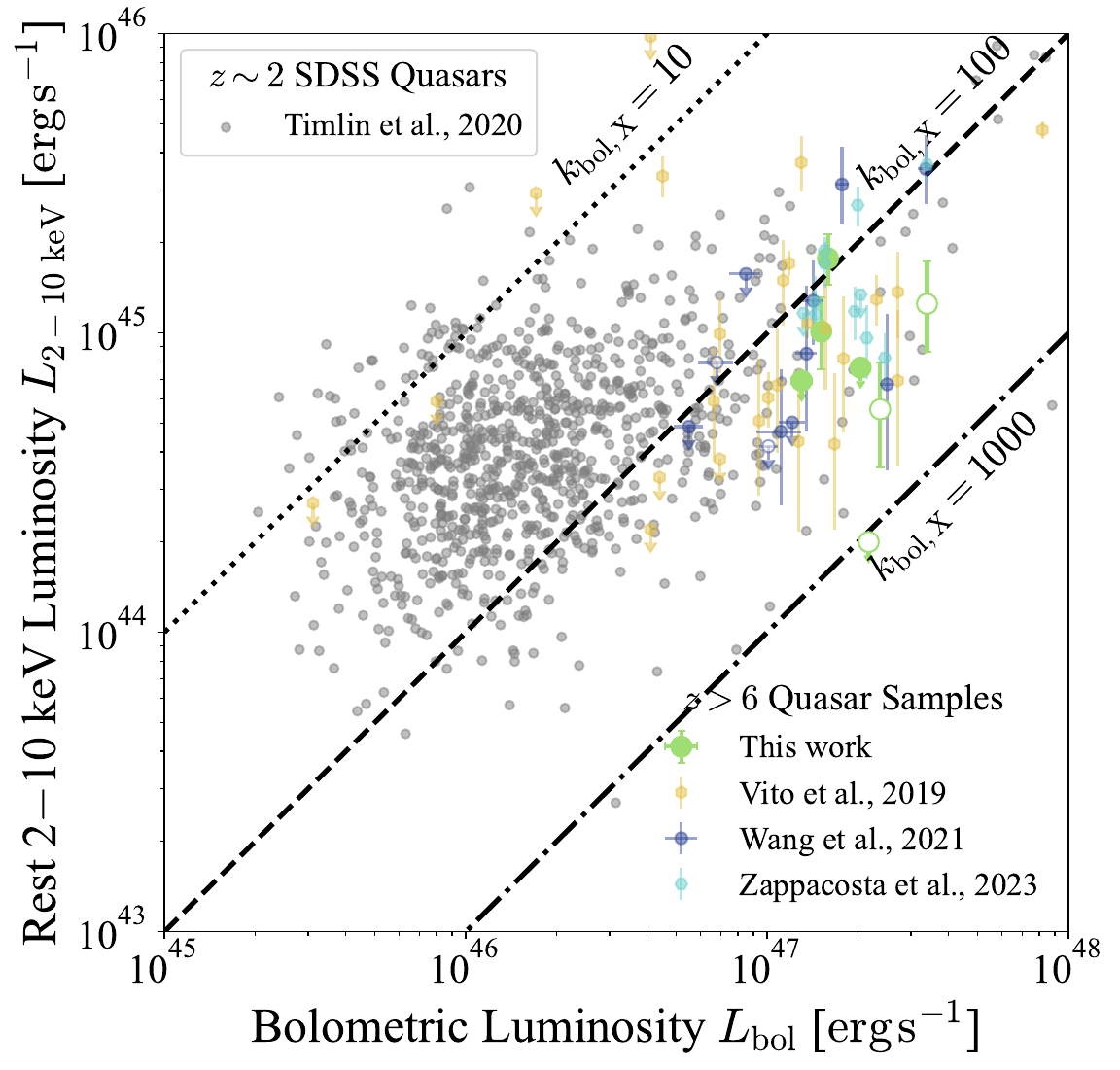}
    \caption{Rest-frame 2$-$10~keV luminosity ($L_{\rm 2-10~keV}$) versus bolometric luminosity ($L_{\rm bol}$) of quasar samples at $z>6$. Identified BAL quasars in this work are displayed in the open symbols. Measurements of $z\sim2$ SDSS quasars are denoted by gray circles \citep{Timlin2020MNRAS,Wang2021ApJ}. The X-ray bolometric corrections $k_{\rm bol, X}$ (defined as $k_{\rm bol, X}=L_{\rm bol}/L_{\rm 2-10~keV}$) of 10, 100, and 1000 are denoted by the dotted, dashed, and the dotted-dashed lines, respectively.}
    \label{fig:k_bol}
\end{figure}

We first investigate the relation between the rest $2-10$~keV X-ray luminosity $L_{\rm 2-10~keV}$ and the bolometric luminosity $L_{\rm bol}$, as shown in Figure \ref{fig:k_bol}. An X-ray bolometric correction $k_{\rm bol, X}$ can be calculated as $k_{\rm bol, X}=L_{\rm bol}/L_{\rm 2-10~keV}$. We plot $k_{\rm bol, X}$ of 10, 100, and 1000 in Figure \ref{fig:k_bol}. $k_{\rm bol, X}$ of most quasars included in this work is within $100-1000$, similar to a $z>6.5$ quasar sample in \citet{Wang2021ApJ} and $z>6$ quasar sample in \citet{Zappacosta2023AA}. However, J0923$+$0402 displays a higher $k_{\rm bol, X}$ than other $z>6$ quasars. Using the existing ground-based NIR spectroscopy, J0923+0402 is identified as a BAL quasar with strong absorption features (a balnicity index $BI_{0}=13670~{\rm km\;\!s^{-1}}$) exhibiting blueward of the \ion{C}{4} emission line \citep{Yang2021ApJ,Bischetti2022Natur}. 
BAL quasars at lower-redshifts have been found to be X-ray weak, although there are debates on whether the X-ray weakness is due to X-ray absorption or intrinsic X-ray deficit \cite[e.g., ][]{Brandt2000ApJ,Luo2013ApJ,Luo2014ApJ,Liu2022ApJ,Veilleux2022ApJ}. Apart from J0923$+$0402, J0038$-$1527, and J0706$+$2921 are also identified as BAL quasars \citep{Yang2021ApJ,Bischetti2022Natur}. However, unlike J0923$+$0402, the other two BAL quasars display a similar $k_{\rm bol, X}$ to non-BAL quasars at $z>6.5$, suggesting that the existence of BAL features might not necessarily be related with the X-ray weakness of quasars at $z>6.5$. Nevertheless, our current sample size is too small to statistically investigate the X-ray properties of $z>6.5$ BAL quasars. 

\subsection{Optical/UV to X-ray Spectral Slope $\alpha_{\rm OX}$ versus Quasar UV Luminosity $L_{\rm 2500\AA}$}

\begin{figure}
    \centering
    \includegraphics[width=0.5\textwidth]{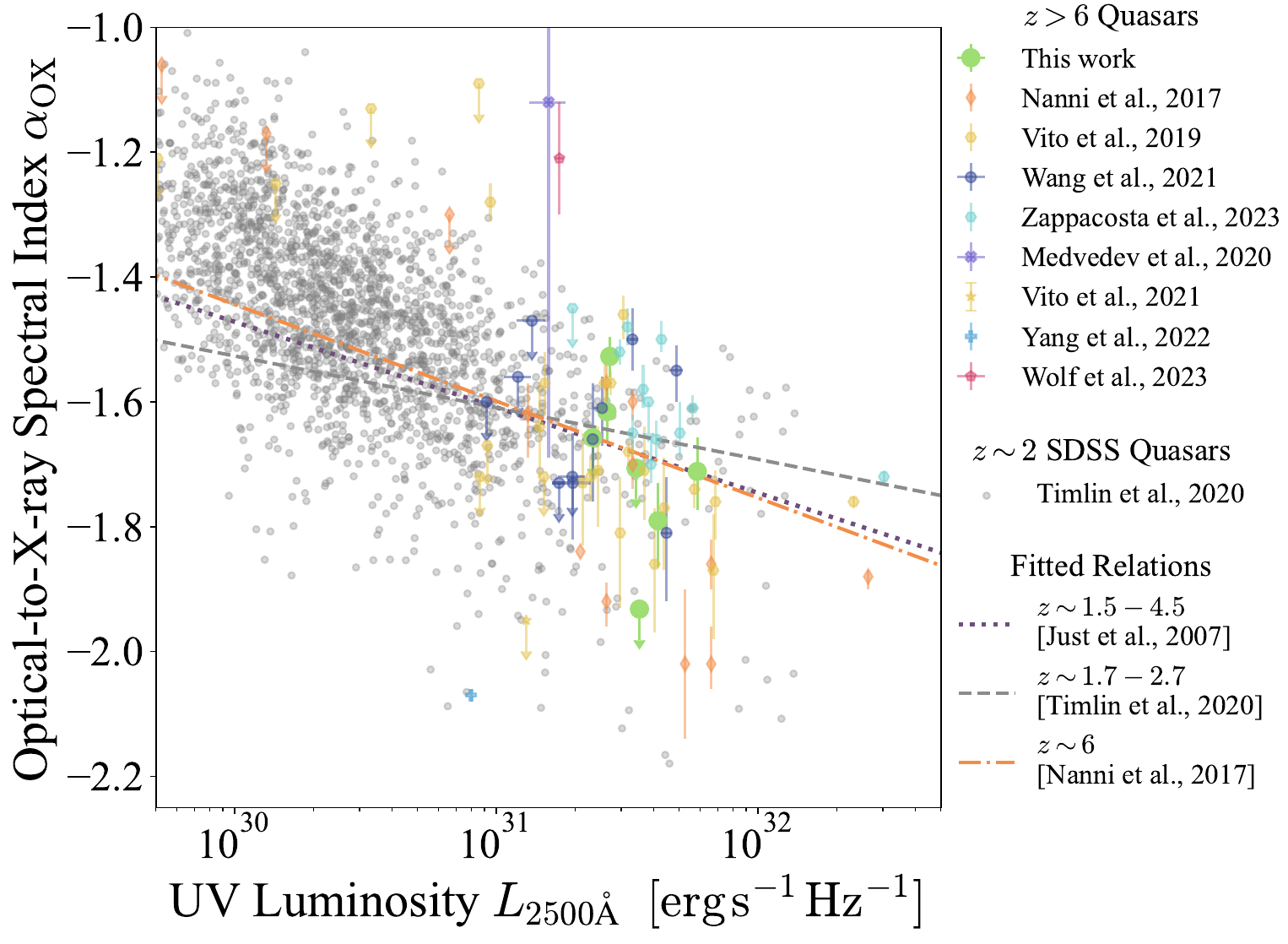}
    \caption{The optical/UV-to-X-ray spectral slope ($\alpha_{\rm OX}$) versus quasar monochromatic UV luminosity at rest 2500$~{\rm \AA}$ ($L_{\rm 2500\AA}$) of quasars at $z>6$. Measurements of $z\sim2$ SDSS quasars are denoted by gray circles \citep{Timlin2020MNRAS} We plot best-fit $\alpha_{\rm OX}-L_{\rm 2500\AA}$ relations of lower-redshift quasars from \citealt{Just2007ApJ} (purple dashed line, $1.5<z<4.5$), \citealt{Timlin2020MNRAS} (green dotted line, $1.7<z<2.7$), and \citealt{Nanni2017AA} (orange dashed-dotted line, $z\sim6$). }
    \label{fig:aox_l2500}
\end{figure}

$\alpha_{\rm OX}$ is a characteristic parameter for the optical/UV-to-X-ray SED shape of quasars, indicating the relative strength of the disk emission (the UV luminosity) to the corona emission (the X-ray luminosity). As shown in previous studies \citep{Nanni2017AA,Vito2019AA,Wang2021ApJ}, $z>6$ quasars display a similar $\alpha_{\rm OX}-L_{\rm 2500\AA}$ relation to $z<6$ quasars \citep{Just2007ApJ,Timlin2020MNRAS}. 
Figure \ref{fig:aox_l2500} shows the relation between $\alpha_{\rm OX}$ and $L_{\rm 2500\AA}$ of $z>6$ quasars. 
Apart from seven $z>6.5$ quasars studied in this work, we also show quasar samples from \citet{Wang2021ApJ} and \citet{Zappacosta2023AA}, and measurements of individual $z>6$ quasars \citep{Medvedev2020MNRAS,Vito2021AA,Yang2022ApJ,Wolf2023AA}. 
Most $z>6$ quasars follow the trends of the $\alpha_{\rm OX}-L_{\rm 2500\AA}$ relation measured from low-redshift quasar samples \citep{Just2007ApJ,Timlin2020MNRAS} and are consistent with the fitted $\alpha_{\rm OX} - L_{\rm 2500\AA}$ relation by including a $z\sim6$ quasar sample \citep{Nanni2017AA}, indicating there is no significant redshift evolution in the $\alpha_{\rm OX}-L_{\rm 2500\AA}$ relation. However, the scatter in the $\alpha_{\rm OX}-L_{\rm 2500\AA}$ relation of $z>6$ quasars is non-negligible. Notably, there are $z>6$ quasars detected by \textit{eROSITA} \cite[e.\,g.,\,][]{Medvedev2020MNRAS,Wolf2023AA}, showing significantly enhanced X-ray emission 
compared with the best-fit $\alpha_{\rm OX} - L_{\rm 2500\AA}$ relations \citep{Just2007ApJ,Nanni2017AA,Timlin2020MNRAS}. 
Such enhanced X-ray emission could be associated with the existence of radio jets 
\cite[e.g.,][]{Belladitta2020AA,Medvedev2021MNRAS}, but some radio-quiet quasars can also display unusual X-ray enhancement \cite[e.g.,][]{Vito2019AA,Wolf2023AA}. 
In addition, a few quasars are observed to be significantly X-ray weaker (e.\,g.,\, J0439$+$1634 included in \citet{Yang2022ApJ}, J0923+0402, and PSOJ167$-$13 in \citet{Vito2021AA}) than the best-fit $\alpha_{\rm OX} - L_{\rm 2500\AA}$ relation of $z\sim6$ quasars \citep{Nanni2017AA}. The X-ray weakness of J0439$+$1634 and J0923+0402 might be associated with the fact that they are both identified as BAL quasars based on the ground-based NIR spectroscopy \citep{Yang2021ApJ}. Although PSOJ167$-$13 is not identified as a BAL quasar based on existing ground-based NIR spectroscopy, it displays a high \ion{C}{4} blueshift of $>4000~{\rm km\;\!s^{-1}}$ \citep{Vito2021AA}, which could be associated with a soft optical/UV-to-X-ray SED shape \cite[e.\,g.,\,][]{Richards2011AJ}. We will discuss the relation between $\alpha_{\rm OX}$ and \ion{C}{4} blueshift of $z>6.5$ quasars in \S\ref{sec:aox_c4}. 



\subsection{$\alpha_{\rm OX}$ versus \ion{C}{4} blueshift velocity} \label{sec:aox_c4}

\begin{figure}
    \centering
    \includegraphics[width=0.45\textwidth]{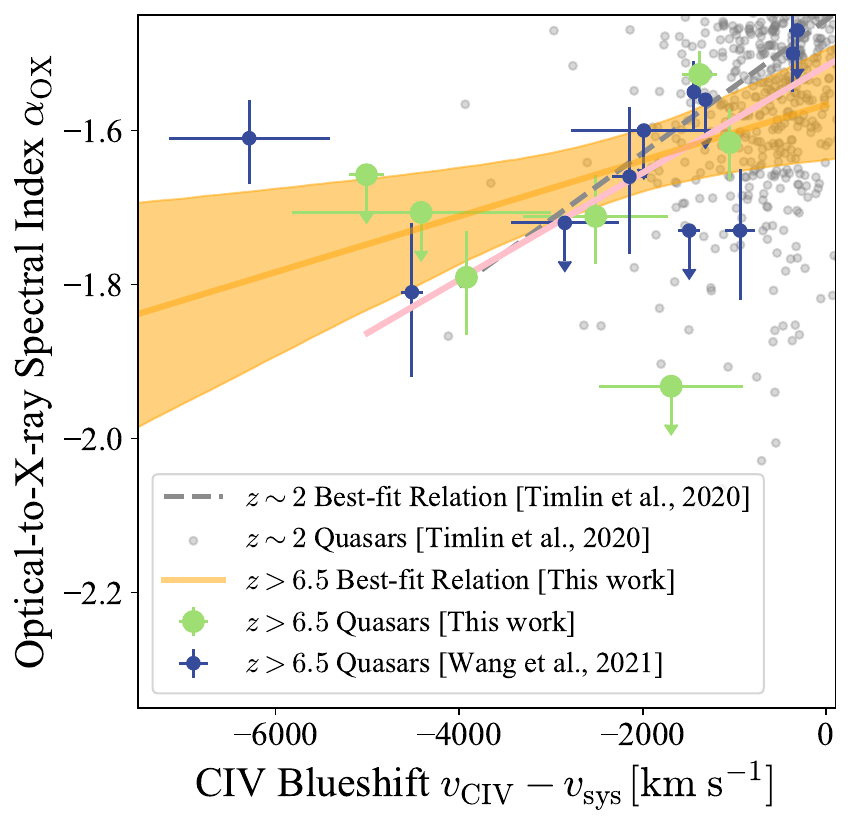}
    \caption{The correlation between optical-to-X-ray spectral index $\alpha_{\rm OX}$ and \ion{C}{4} blueshift velocity. Measurements of $z\sim2$ SDSS quasars are denoted by gray circles, and the best-fit $\alpha_{\rm OX}-$\ion{C}{4} blueshift relation of $z\sim2$ SDSS quasars is shown by the gray dashed line \citep{Timlin2020MNRAS}. The best-fit $\alpha_{\rm OX}-$\ion{C}{4} blueshift relation at $z>6.5$ and its 1$\sigma$ confidence intervals are shown by the orange solid line and the orange shaded region. The best-fit $\alpha_{\rm OX}-$\ion{C}{4} blueshift relation at $z>6.5$ by excluding J1342$+$0928 and J0923$+$0402 is shown by the pink solid line.}
    \label{fig:aox_v_offset}
\end{figure}

Reionization-era quasars are observed to exhibit a higher average \ion{C}{4} blueshift than luminosity-matched quasar samples at lower-redshifts \citep{Mazzucchelli2017ApJ,Meyer2019MNRAS,Schindler2020ApJ,Yang2021ApJ}, suggesting the existence of prominent disk winds \citep{MC1995ApJ,Richards2012ASPC}. While for $z\sim2$ quasars showing a higher \ion{C}{4} blueshift, they often display a softer $\alpha_{\rm OX}$ (i.e., a softer optical/UV-to-X-ray SED shape) \citep{Richards2011AJ,Timlin2020MNRAS}. To investigate whether $z>6.5$ quasars display a correlation between $\alpha_{\rm OX}$ and \ion{C}{4} blueshift velocity, we calculate the \ion{C}{4} blueshift velocity, using $v_{\rm CIV}-v_{\rm sys}=(z_{\rm CIV}-z_{\rm sys})/(1+z_{\rm sys})\times c$, where the $z_{\rm sys}$ is the quasar systemic redshift. Although the \ion{Mg}{2} emission line can represent the systemic redshift of SDSS quasars on average \citep{Shen2016ApJ}, it can significantly deviate the systemic redshift of individual quasars \cite[e.g.,][]{Richards2002AJ}. Moreover, reionization-era quasars are found to show an average \ion{Mg}{2} blueshift of $\sim400~{\rm km\;\!s^{-1}}$, compared with the [\ion{C}{2}] 158 \textmu m emission line \citep{Schindler2020ApJ}, which is much higher than the \ion{Mg}{2} blueshift of $57~{\rm km\;\!s^{-1}}$ (compared with \ion{Ca}{2}) measured from $0.1<z<4.5$ SDSS-Reverberation Mapping quasars \citep{Shen2016ApJ}.  
All quasars in this work and \citet{Wang2021ApJ} have been observed by ALMA, therefore, we thus use the redshift of [\ion{C}{2}] 158 \textmu m ($z_{\rm [C II]}$) as the systemic redshift of quasars \citep{Decarli2018ApJ,Banados2019ApJ,Venemans2019ApJ,Venemans2020ApJ,Yang2021ApJ,Wang2024ApJ}. 
Figure \ref{fig:aox_v_offset} shows the correlation between $\alpha_{\rm OX}$ and the CIV blueshift velocity. 
As addressed in \citet{Wang2021ApJ}, J1342$+$0928 could be an outlier, well above the $\alpha_{\rm OX}-$\ion{C}{4} blueshift relation reported in \citet{Timlin2020MNRAS}. It is worthy noting that none of quasars in \citet{Timlin2020MNRAS} display such a high \ion{C}{4} blueshift as J1342$+$0928 (see Figure \ref{fig:aox_v_offset}), 
therefore, it remains unclear whether the $\alpha_{\rm OX}$–\ion{C}{4} blueshift relation measured in \citet{Timlin2020MNRAS} can be extrapolated to a \ion{C}{4} blueshift of 6000~${\rm km\;\!s^{-1}}$. 
Apart from J1342$+$0928, the BAL quasar J0923+0402 is another potential outlier to the $\alpha_{\rm OX}-$\ion{C}{4} blueshift relation in \citet{Timlin2020MNRAS}, displaying a much weaker X-ray emission at a given \ion{C}{4} blueshift velocity. \citet{Timlin2020MNRAS} excluded BAL quasars from their analysis, because BAL quasars can potentially show X-ray weakness due to additional absorption. However, as shown in Figures \ref{fig:k_bol} and \ref{fig:aox_l2500}, the BAL quasars at $z>6.5$ do not uniformly show X-ray weakness. Given the limited number of quasars with X-ray observations at these redshifts, we include BAL quasars in our analysis in order to examine the correlation between $\alpha_{\rm OX}$ and \ion{C}{4} blueshift velocity. Nevertheless, a larger sample will be needed to access the influence of BAL quasars on this correlation. 

We calculate the Spearman rank correlation coefficient between $\alpha_{\rm OX}$ and \ion{C}{4} blueshift velocity using ASURV package \citep{Feigelson1985ApJ,Isobe1986ApJ,Lavalley1992ASPC}, which can treat upper limits of $\alpha_{\rm OX}$. We obtain the corresponding Spearman $\rho=0.38$ with a $p$-value$=0.12$. We also calculate the Spearman coefficient by excluding J1342+0928 and J0923$+$0402 from the sample, and the corresponding Spearman $\rho$ is $0.59$ with a $p$-value$=0.02$. This suggests a potential positive correlation between $\alpha_{\rm OX}$ and CIV blueshift velocity for quasars at $z>6.5$, similar to $z\sim2$ quasars \citep{Timlin2020MNRAS}, but the current sample size of quasars at $z>6.5$ is still limited to fully probe this correlation.
We then perform a linear fitting to the relation between $\alpha_{\rm OX}$ and CIV blueshift using a Bayesian approach \texttt{linmix}\footnote{The python version of linmix is used in our analysis: https://linmix.readthedocs.io/en/latest/index.html} \citep{Kelly2007ApJ}, which performs linear regression with measurement errors and censored data (upper limits or lower limits) in the datasets. 
The fitted relation from \texttt{linmix} between $\alpha_{\rm OX}$ and CIV blueshift $v_{\rm CIV}-v_{\rm sys}$ is: $\alpha_{\rm OX}=(3.63\pm2.90)\times10^{-5}\times[\frac{v_{\rm CIV}-v_{\rm sys}}{\rm km\;\!s^{-1}}]+(-1.57\pm0.08)$\footnote{The best-fit relation by excluding J1342$+$0928 and J0923+0402 is: $\alpha_{\rm OX}=(7.07\pm3.14)\times10^{-5}\times[\frac{v_{\rm CIV}-v_{\rm sys}}{\rm km\;\!s^{-1}}]+(-1.51\pm0.07)$ (the pink solid line in Figure \ref{fig:aox_v_offset})}, and is shown in the orange solid line in Figure \ref{fig:aox_v_offset}. The $1\sigma$ confidence intervals of the relation are denoted by the orange shaded region in Figure \ref{fig:aox_v_offset}. 

The correlation between $\alpha_{\rm OX}$ and \ion{C}{4} blueshift suggests the quasar optical/UV-to-X-ray SED shape is critical for launching disk winds, which could be traced by the \ion{C}{4} blueshift \citep{Richards2012ASPC,Matthews2023MNRAS}. In the radiation-driven disk wind model, UV photons emitted by the accretion disk would accelerate the disk wind through resonance lines such as Lyman $\alpha$ and \ion{C}{4}, 
while X-ray emission can cause the over-ionization of the gas, reducing the opacity, thus decreasing the wind strength \citep{MC1995ApJ,Proga2000ApJ,Proga2003ApJ}, therefore, a soft SED shape is most effective for driving disk winds \cite[e.g.,][]{Rivera2022ApJ}. Compared with the $\alpha_{\rm OX}$ and \ion{C}{4} blueshift relation of $z\sim2$ quasars in \citet{Timlin2020MNRAS}, $z>6.5$ quasars follow a very similar relation. Although the \ion{C}{4} blueshift in \citet{Timlin2020MNRAS} is measured through the velocity offset between \ion{C}{4} and \ion{Mg}{2}, considering the intrinsic uncertainty of $\sim200~{\rm km\;\!s^{-1}}$ when using the \ion{Mg}{2} redshift as systemic redshift, measured from SDSS Reverberation Mapping quasars \citep{Shen2016ApJ}, the best-fit $\alpha_{\rm OX}$ $-$ \ion{C}{4} blueshift relation of $z\sim6.5$ quasars is consistent within $1\sigma$ with the best-fit $\alpha_{\rm OX}$ $-$ \ion{C}{4} blueshift relation at $z\sim2$ \citep{Timlin2020MNRAS}. This indicates no significant redshift evolution in 
the mechanism for quasar radiation driven winds. We emphasize that the correlation between $\alpha_{\rm OX}$ and \ion{C}{4} blueshift is tentative for quasars at $z>6.5$, and a larger sample is needed to fully probe this correlation and to assess the influence of BAL quasars on this correlation.

\subsection{X-ray Spectral Analysis of $z>6.5$ Quasars}
\begin{figure}
    \centering
    \includegraphics[width=0.5\textwidth]{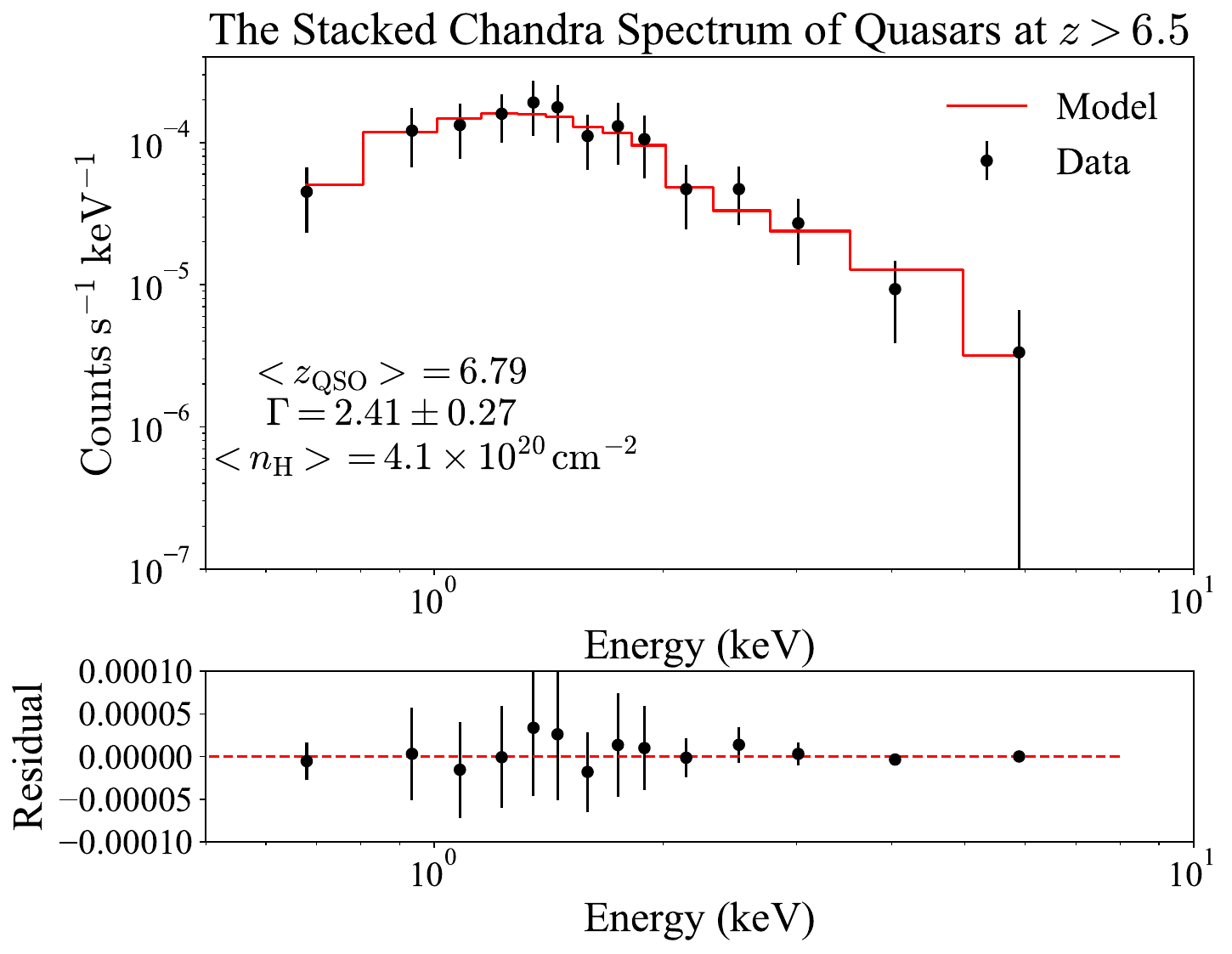}
    \caption{\textit{Top} -- The stacked spectrum of 11 quasars at $z>6.5$ detected by \textit{Chandra} in 10 counts per bin (black). In total, there are 122 net counts in 0.5--7~keV. The best-fit absorbed power-law model is denoted by the red line with a photon index $\Gamma$ of $2.41\pm0.27$. \textit{Bottom} -- The residual data (data-model).}
    \label{fig:stacked_xray_spectrum}
\end{figure}

\begin{figure}
    \centering
    \includegraphics[width=0.45\textwidth]{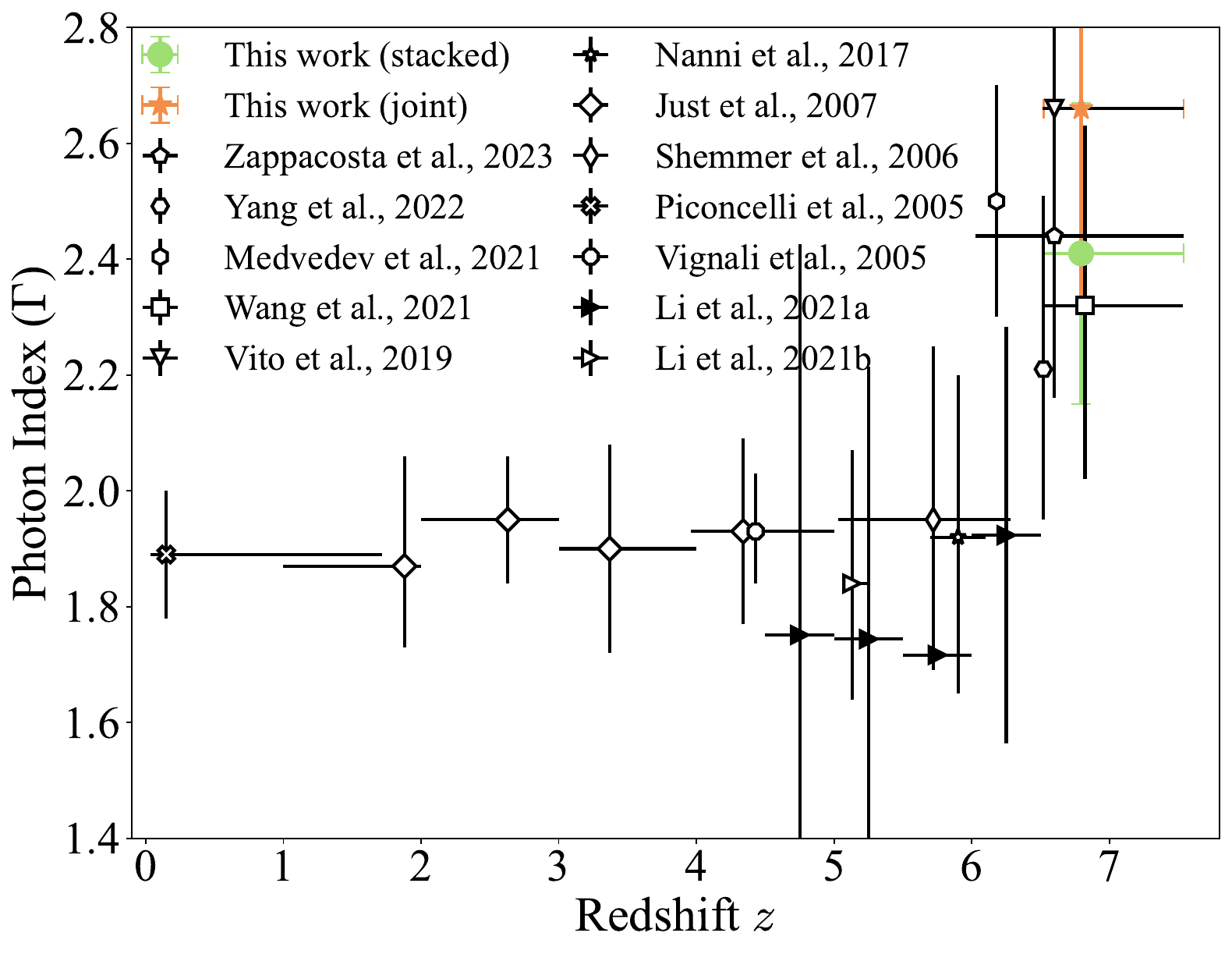}
    \caption{The photon index $\Gamma$ of luminous quasars as a function of redshift $z$.}
    \label{fig:photon_index_z_evolution}
\end{figure}

For quasars detected in $0.5-7$~keV by \texttt{wavdetect}, the net $0.5-7$~keV counts range from 8 to 34, preventing tight constraints on X-ray spectral properties using individual sources. To enable X-ray spectral analysis using a limited number of X-ray counts we combine the X-ray spectrum of all $z>6.5$ quasars and then perform the spectral fitting of the stacked spectrum.

To obtain a stacked X-ray spectrum, we first employ \texttt{combine\_spectra} to combine the source and the background spectra of all $z > 6.5$ quasars detected by Chandra. 
The binomial probability $P_{\rm B}$ of J0252$-$0503 and J1007$+$2115 indicates the detection is tentative in $0.5-7$~keV, and therefore we exclude them from the spectral analysis. To increase the number of X-ray counts used in the spectral analysis, when combing the spectra, we also include Chandra observations of $z>6.5$ quasars (J0224$-$4711, J0226$+$0302, J1120$+$0641, J1342$+$0928, J2132$+$1217, J2232$+$2930) from \citet{Wang2021ApJ} and J0921$+$0007 from \citet{Wolf2023AA}. 
The stacked spectrum of 11 quasars at $z > 6.5$ contains a total of 122 net counts in the 0.5--7~keV band. 
For the stacked spectrum, we use Sherpa to group the 0.5$-$7~keV counts into 10 counts per bin. We fit the stacked spectrum with a redshifted power-law model plus a galactic absorption (\texttt{xsphabs*xszpowerlw}). We adopt an average galactic hydrogen column density $\langle n_{\rm H} \rangle =4.1\times10^{20}\;\!{\rm cm^{-2}}$ around these 11 quasar sightlines as the absorbed column density \citep{HI4PI2016AA}, and the average quasar [\ion{C}{2}] redshift $\langle z_{\rm QSO}\rangle=6.79$ as the source redshift. We fix the absorbed column density and the source redshift when fitting the stacked X-ray spectrum, and we obtain the best-fit $\Gamma=2.41\pm0.27$. In Figure \ref{fig:stacked_xray_spectrum}, we plot the stacked X-ray spectrum of 11 $z>6.5$ quasars in black, and the best-fit absorbed redshifted power-law model in red. 

To avoid the potential bias caused by bright X-ray sources in the stacked spectrum, we also perform a joint analysis of the X-ray spectra of all 11 quasars. Due to a limited number of X-ray counts in each X-ray spectrum, we employ \texttt{wstat} in our analysis, which uses Cash statistics \citep{Cash1979ApJ}, and can also treat associated background spectra with Poisson statistics. For each quasar spectrum, we assume a redshifted power-law model with a galactic absorption (\texttt{xsphabs*xszpowerlw}), and fix the galactic absorption and source redshift. We assume all X-ray spectra have the same photon index, {allow the normalization factor of each spectrum to vary}, and jointly fit all 11 X-ray spectra. The best-fit $\Gamma$ from the joint spectral analysis of 11 quasars is $2.67\pm0.34$, which is in agreement with the $\Gamma=2.41\pm0.27$ measured from the stacked X-ray spectrum. 

In Figure \ref{fig:photon_index_z_evolution}, we present the redshift evolution of the photon index $\Gamma$, measured from the X-ray spectra of luminous quasars. 
Here we only include $\Gamma$ derived from individual/stacked spectra with total net counts $>50$. At $z\lesssim6$, previous studies show a $\Gamma\sim1.8-1.9$ without an evident redshift evolution \citep{Piconcelli2005AA,Vignali2005AJ,Shemmer2006ApJ,Just2007ApJ,Nanni2017AA,Li2021MNRAS,Li2021ApJa}. At $z>6$, spectral analysis of quasar samples \citep{Vito2019AA,Wang2021ApJ,Zappacosta2023AA} and spectral analysis of individual quasars \cite[with $>300$ net counts,][]{Medvedev2021MNRAS,Yang2022ApJ} both show significantly steeper $\Gamma$ than $z<6$ quasars. $\Gamma$ measured from our analysis of 11 $z>6.5$ quasars is consistent with those previous studies of $z>6$ quasars, and is also significantly higher than $z<6$ quasars, reinforcing the steepening of photon index for $z>6$ quasars.


\begin{figure}
    \centering
    \includegraphics[width=0.45\textwidth]{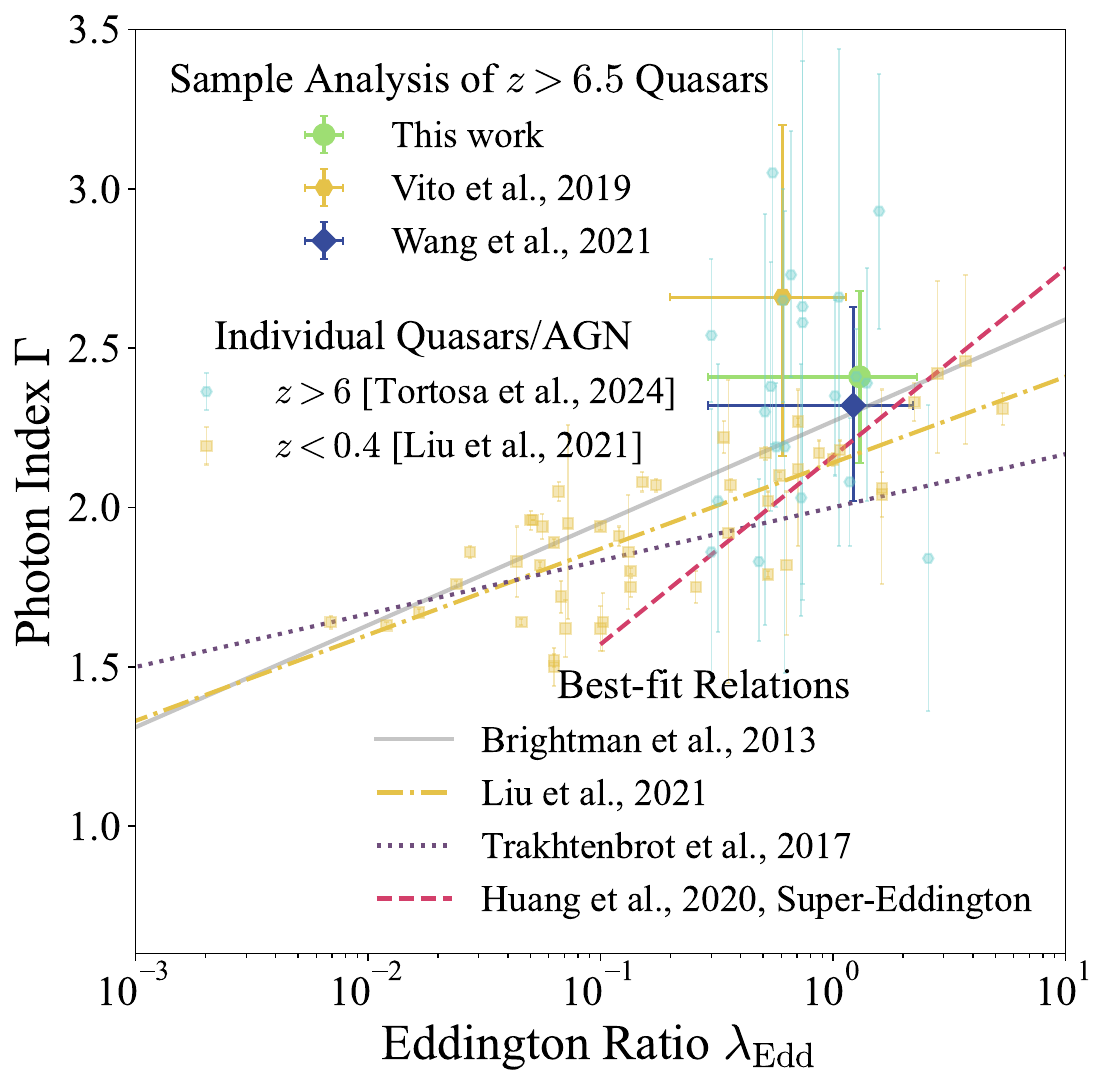}
    \caption{The photon index $\Gamma$ as a function of the Eddington ratio $\lambda_{\rm Edd}$. The average photon index of quasar samples at $z>6.5$ are denoted by the green circle (this work), yellow hexagon \citep{Vito2019AA}, and blue diamond \citep{Wang2021ApJ}, with error bars of the average Eddington ratio representing the range of the quasar Eddington ratio distribution. We also show individual $\Gamma$ measurements of $z>6$ quasars (teal hexagons, \citealt{Tortosa2024AA}) and $z<0.4$ AGN (yellow squares, \citealp{Liu2021ApJ}). The best-fit $\Gamma-\lambda_{\rm Edd}$ relations from \citet{Brightman2013MNRAS,Liu2021ApJ,Trakhtenbrot2017MNRAS} and \citet{Huang2020ApJ} are plotted in the gray solid, yellow dashed-dotted, purple dotted, and pink dashed lines, respectively.}
    \label{fig:gamma_versus_lambda_Edd}
\end{figure}

A few hypotheses have been proposed to explain the steepening in the X-ray spectrum of $z>6$ quasars, including: (1) redshift evolution in the disk-corona coupling or coronal properties for $z>6$ quasars leads to a higher $\Gamma$ \citep{Zappacosta2023AA}; or (2) the high accretion rate of $z>6.5$ quasars results in a steeper $\Gamma$ \citep{Wang2021ApJ}. For the latter, a statistically significant positive corelation between $\Gamma$ and Eddington ratio ($\lambda_{\rm Edd}$) has been found from $z<3$ quasars \citep{Shemmer2008ApJ,Risaliti2009ApJ,Brightman2013MNRAS}. This suggests the coronal cooling becomes more efficient
at a high $\lambda_{\rm Edd}$, reducing the corona temperature, resulting in a softer X-ray spectrum associated with a higher $\Gamma$ \citep{Pounds1995MNRAS}. In Figure \ref{fig:gamma_versus_lambda_Edd}, we show the average Eddington ratio and $\Gamma$ of high-redshift quasar samples from joint spectral analysis (this work and \citealt{Wang2021ApJ}), as well as individual $\Gamma$ measurements of $z>6$ quasars \citep{Tortosa2024AA} and $z<0.4$ AGN \citep{Liu2021ApJ}. We also show the best-fit $\Gamma-\lambda_{\rm Edd}$ relations in \citet{Brightman2013MNRAS,Trakhtenbrot2017MNRAS,Huang2020ApJ} and \citet{Liu2021ApJ}. \citet{Tortosa2024AA} report no correlation between $\Gamma$ and $\lambda_{\rm Edd}$ of $z>6$ quasars. However, the depth of most X-ray observations can only moderately constrain $\Gamma$ of individual $z>6$ quasars, leaving large uncertainties in $\Gamma$. Moreover, the dynamic range of the $\lambda_{\rm Edd}$ distribution of current $z>6$ quasar sample is very limited \cite[e.g.,][]{Schindler2020ApJ,Yang2021ApJ}, posing a challenge to constrain the $\Gamma-\lambda_{\rm Edd}$ relation using $z>6$ quasars alone. We therefore compare the $\Gamma$ derived from $z>6.5$ quasar samples with the $\Gamma-\lambda_{\rm Edd}$ relation measured from lower-redshift samples \citep{Brightman2013MNRAS,Trakhtenbrot2017MNRAS,Liu2021ApJ}. We note that the $\Gamma-\lambda_{\rm Edd}$ relations measured from \citet{Brightman2013MNRAS,Trakhtenbrot2017MNRAS} and \citet{Liu2021ApJ} all show a positive correlation, though the slopes differ substantially. We find the average high $\Gamma$ of $z>6.5$ quasars are in good agreement with the $\Gamma-\lambda_{\rm Edd}$ relation reported in \citet{Brightman2013MNRAS} and \citet{Liu2021ApJ} within $1\sigma$, but in a weak tension with the flatter $\Gamma-\lambda_{\rm Edd}$ relation from \citet{Trakhtenbrot2017MNRAS}. \citet{Trakhtenbrot2017MNRAS} and \citet{Huang2020ApJ} have discussed several factors contributing to the scatter in the $\Gamma-\lambda_{\rm Edd}$ relation, including uncertainties in single-epoch black hole mass estimates (both systematic errors and measurement errors), differences in X-ray spectral models, the choice of the energy band for spectral fitting, and variations in accretion processes at different accretion rates. 
From a sample of SDSS quasars at $z<0.7$, \citet{Huang2020ApJ} identify those quasars with a dimensionless accretion rate $\mathscr{\dot{M}}>3$ as super-Eddington quasars. The dimensionless accretion rate $\mathscr{\dot{M}}$ is defined as $\mathscr{\dot{M}}=\dot{M}c^2/L_{\rm Edd}=\lambda_{\rm Edd}/\eta$, where $\dot{M}$ is mass accretion rate, $L_{\rm Edd}$ is the Eddington luminosity, $\lambda_{\rm Edd}$ is the Eddington ratio, and $\eta$ is the radiative efficiency. $\mathscr{\dot{M}}>3$ is corresponding to $\lambda_{\rm Edd}>0.3$ when adopting $\eta=0.1$. \citet{Huang2020ApJ} report that the slope of the $\Gamma-\lambda_{\rm Edd}$ relation is steeper for super-Eddington quasars than sub-Eddington quasars, suggesting that disk-corona connections could be different for super-Eddington quasars. 
Given the high $\lambda_{\rm Edd}$ of quasars at $z>6.5$, in Figure \ref{fig:gamma_versus_lambda_Edd}, we show the best-fit $\Gamma-\lambda_{\rm Edd}$ relation of super-Eddington quasars in \citet{Huang2020ApJ}, and find the average $\Gamma$ and $\lambda_{\rm Edd}$ of luminous $z>6.5$ quasars is very consistent with the $\Gamma-\lambda_{\rm Edd}$ relation of super-Eddington quasars at $z<0.7$ \citep{Huang2020ApJ}. 
This suggests the steep $\Gamma$ of $z>6.5$ quasars is mainly driven by high Eddington ratios, instead of an evolution in disk-corona connections across cosmic time. 



\subsection{Infrared Luminosity versus X-ray Luminosity}  

\begin{figure*}
    \centering
    \includegraphics[width=0.9\textwidth]{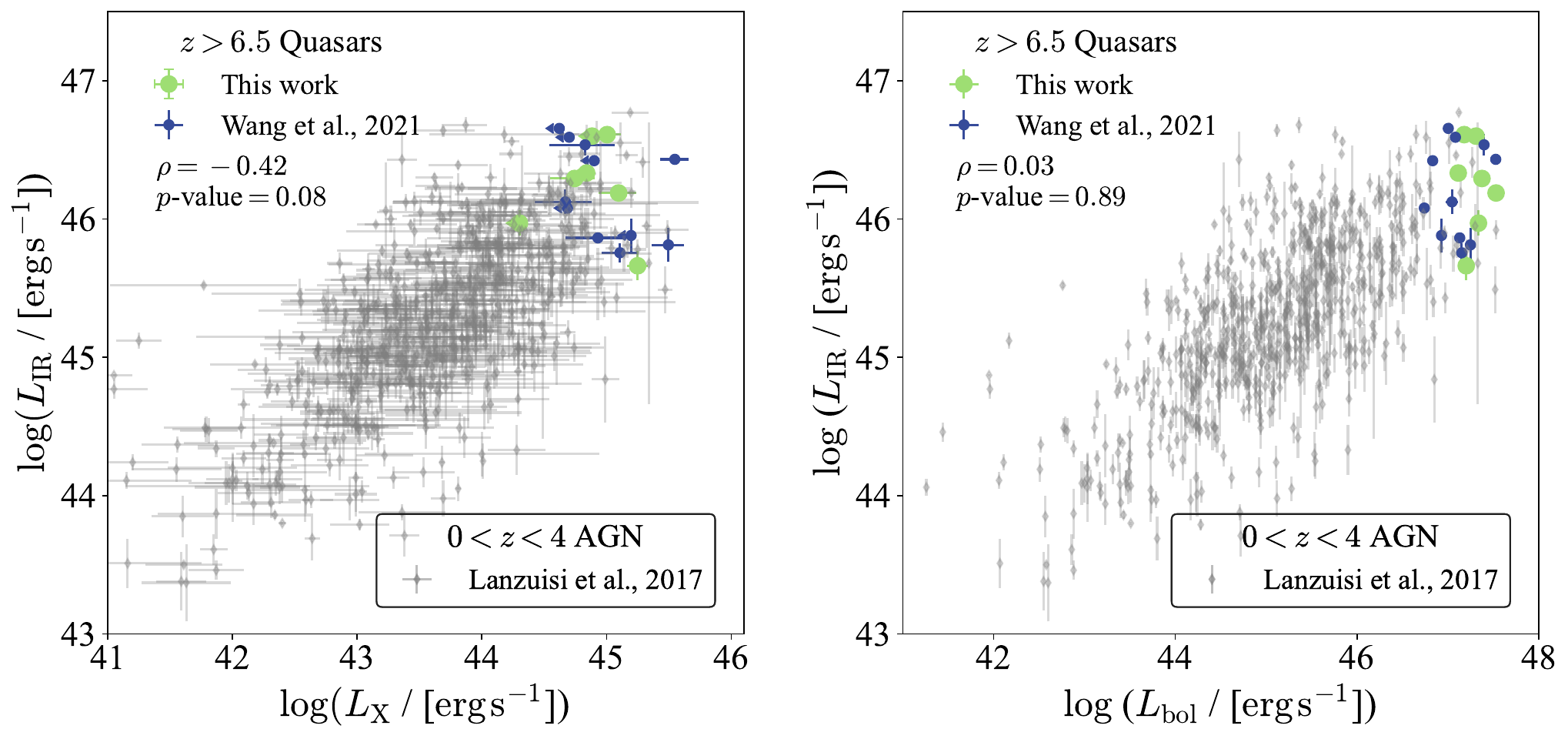}
    \caption{\textit{Left} -- Infrared luminosity $L_{\rm IR}$ versus rest 2$-$10~keV X-ray luminosity $L_{\rm X}$; \textit{Right} --  Infrared luminosity $L_{\rm IR}$ versus bolometric luminosity $L_{\rm bol}$. Measurements of quasars at $z>6.5$ are denoted by green circles (this work), and blue circles \citep{Wang2021ApJ}. Measurements of AGN at $z<4$ are shown by gray diamonds \citep{Lanzuisi2017AA}.}
    \label{fig:lir_lx}
\end{figure*}

Observed tight scaling relations between nearby central SMBHs and their host galaxies \cite[e.g., $M_{\rm BH}-\sigma$ and $M_{\rm BH}-M_{*}$ relations,][]{Magorrian1998AJ,FM2000ApJ,Gebhardt2000ApJ,MH2003ApJ,RV2015ApJ} indicate the co-evolution between SMBHs and their host galaxies (for a review, see \citealp{KH2013ARAA}), implying significant influence of AGN in regulating the star formation activity of the host galaxy. 
It remains debated whether the scaling relations between the central black hole and the host galaxy properties evolve with redshift (e.\,g.,\,\citealp{Peng2006ApJ,Merloni2010ApJ,Zhang2023ApJ}, see also \citealp{SS2013ApJ,Ding2020ApJ,Suh2020ApJ,Sun2025arXiv,Sun2025ApJ}). 
At $z>6$, the central SMBH of the most luminous quasars is often found to be ``overmassive" compared to their host galaxy than the local relation (e.g., \citealp{Stone2023ApJ,Stone2024ApJ,Yue2024ApJ,Onoue2024arXiv}, but see also \citealp{Ding2023Natur,Zhang2023MNRAS,Li2025arXiv,Li2025ApJ,Sun2025arXiv}), suggesting the SMBH growth of $z>6$ quasars is faster than the growth of the host galaxy. 
Through JWST NIRCam and NIRSpec integral field spectroscopy observations, a few $z>6.5$ quasars show a highly blueshifted  [\ion{O}{3}] emission line (e.\,g.,\,\citet{Marshall2023AA,Yang2023ApJb,Decarli2024AA,Liu2024ApJ}, see also \citet{Lyu2024arXiv}), likely associated with the existence of fast, galactic scale outflows \citep{Mullaney2013MNRAS,Carniani2015AA}, indicating a potential impact from the quasar in repelling gas reservoir of the host galaxy \cite[e.g.,][]{Liu2024ApJ,Zhu2025arXiv}. 
In addition, with JWST/NIRSpec observations, the host galaxy of two $z\sim6$ quasars from Subaru High-$z$ Exploration of Low-Luminosity Quasars program \cite[SHELLQs,][]{Matsuoka2016ApJ,Matsuoka2018ApJS,Matsuoka2018PASJ} has displayed post-starburst signatures \citep{Onoue2024arXiv}, implying a critical role of quasars in influencing the star formation of their host galaxy. However, the statistical investigation of the correlation between the black hole accretion rate and the star formation rate is very limited at $z>6$ 
\cite[e.g.,][]{Venemans2018ApJ,Venemans2020ApJ,Wang2021ApJ,Wang2024ApJ}.

To investigate the relation between SMBH growth and the host galaxy growth at $z\sim6$, we use $L_{\rm X}$ and $L_{\rm bol}$ as indicators of the SMBH accretion rate, and study their correlations with host galaxy star formation rate measured from the infrared luminosity \cite[e.g.,][]{Kennicutt1998ARAA,Calzetti2013seg}. We adopt the infrared luminosity ($L_{\rm IR}$) estimated from dust continuum flux measured from ALMA observations integrated over the rest-frame $8-1000\;\!{\rm \mu}$m \citep{Wang2024ApJ}, assuming the Haro 11 template. Haro 11 is a compact, moderately metal-poor galaxy, considered analogous to high-redshift quasar host galaxies \citep{Lyu2016ApJ}. The star formation rate can be calculated based on $L_{\rm IR}$, using ${\rm SFR/( M_{\odot}\;\!yr^{-1}})=5\times10^{-44}\;\!L_{\rm IR}/({\rm erg\;\!s^{-1}})$ \citep{Lyu2016ApJ}. 


Figure \ref{fig:lir_lx} shows the correlations between ${\rm log} (L_{\rm IR})$ and ${\rm log} (L_{\rm X})$ (left), and ${\rm log} (L_{\rm IR})$ and ${\rm log} (L_{\rm bol})$ (right), with corresponding Spearman correlation coefficients of $-0.42$ ($p$-value=0.08) and 0.03 ($p$-value=0.89) for quasars at $z>6.5$, calculated by ASURV. 
The Spearman correlation coefficients 
(between $L_{\rm IR}$ and $L_{\rm X}$ or either $L_{\rm IR}$ and $L_{\rm bol}$) indicate no significant correlations between the star formation luminosity of the quasar host galaxy and the quasar luminosity. 
This result is consistent with previous observational studies of quasars at $z>6.5$ \citep{Venemans2018ApJ,Venemans2020ApJ,Wang2021ApJ} and simulations \citep{Valentini2021MNRAS}. This would suggest for luminous quasars at $z>6.5$, the quasar luminosity is not correlated to the star formation luminosity in the same way as low-redshift relations \cite[e.g.,][]{Netzer2009MNRAS,Page2012Natur,Rosario2012AA,Xu2015ApJ,Lanzuisi2017AA}. 
We emphasize that no correlations between $L_{\rm IR}$ and $L_{\rm X}$ or $L_{\rm IR}$ and $L_{\rm bol}$ in our analysis do not imply that quasars do not regulate the star formation activity at $z>6.5$. Because we only include the most luminous $z>6.5$ quasars in our analysis, which can only represent the most extreme black holes during the epoch of reionization, the absence of an observed correlation can be caused by the narrow dynamic ranges in $L_{\rm IR}$, $L_{\rm X}$ and $L_{\rm bol}$. Compared with the AGN sample at $0<z<4$ in \citet{Lanzuisi2017AA}, our quasar sample at $z>6.5$ occupies the bright-end of $L_{\rm IR}$, $L_{\rm X}$ and $L_{\rm bol}$. As reported in \citet{Wang2011AJ}, the slope of ${\rm log}(L_{\rm FIR})-{\rm log}(L_{\rm bol})$ is 0.62 for $z\sim2-6$ quasars. If the ${\rm log}(L_{\rm IR})-{\rm log}(L_{\rm bol})$ and ${\rm log}(L_{\rm IR})-{\rm log}(L_{\rm X})$ relations follow a similar flat slope as the ${\rm log}(L_{\rm FIR})-{\rm log}(L_{\rm bol})$ relation, a broader dynamic range in $L_{\rm IR}$, $L_{\rm X}$ and $L_{\rm bol}$ will be needed to test the correlation between star formation luminosity and quasar accretion luminosity. With fainter quasar/AGN samples from Subaru \citep{Matsuoka2016ApJ,Matsuoka2018ApJS,Matsuoka2018PASJ}, JWST \cite[e.g.,][]{Matthee2024ApJ,Maiolino2024AA,Lin2024ApJ}, Euclid \cite[e.g.,][]{Euclid2019AA}, Rubin and Roman \cite[e.g.,][]{Tee2023ApJ}, the relation between SMBH growth and host galaxy star formation will be investigated statistically over a broad dynamic range. 
In addition, {$L_{\rm X}$ and $L_{\rm bol}$ trace the AGN activity on a black hole accretion timescale much shorter than the star formation timescale ($\gtrsim10-100$~Myr) traced by $L_{\rm IR}$ \citep{Kennicutt1998ARAA,Hickox2012MNRAS,Calzetti2013seg}. While from the quasar proximity zone measurements of quasars at $z>6$, the quasar lifetime, recording the duration of most recent quasar accretion activity, is often found to be less than $10^{5-6}~{\rm yr}$ \citep{Eilers2017ApJ,Eilers2021ApJ,Eilers2025ApJ,Morey2021ApJ}. Some quasars could also display extreme variability on short timescales of $\lesssim10$~yr (e.g., \citet{LaMassa2015ApJ,Ruan2019ApJ,Jin2021ApJ,Gilbert2025arXiv}, but see also \citet{Shen2021ApJ}). This discrepancy between the black hole accretion timescale and the star formation timescale may lead to the absence of an observed correlation between quasar accretion luminosity and star formation luminosity \cite[e.g.,][]{Venemans2018ApJ}.} The conversion factor between $L_{\rm IR}$ and star formation rate also depends on the assumption of the star formation history \cite[e.g.,][]{Calzetti2013seg} and the contribution of AGN to far-infrared emission \cite[e.g.,][]{Dale2012ApJ,Lyu2017ApJ,Lyu2022Univ,Symeonidis2016MNRAS}. Future ALMA and JWST/MIRI observations will further constrain the dust continuum and offer {deeper insights} into the star formation in quasar host galaxies.

\section{Summary}


In this paper, we present new Chandra observations of seven luminous quasars at $z>6.5$. We analyze quasar properties measured from Chandra, ground-based NIR spectroscopy, and ALMA observations. We find: 

\begin{itemize}
    \item $z>6$ quasars generally follow a similar $\alpha_{\rm OX}-L_{\rm 2500\AA}$ relation as $z\sim2$ quasars, suggesting no significant redshift evolution in the optical/UV-to-X-ray SED shape at a given $L_{\rm 2500\AA}$. The scatter of $\alpha_{\rm OX}-L_{\rm 2500\AA}$ relation is significant for $z>6$ quasars. Quasars displaying signatures of fast nuclear winds (e.g., strong broad-absorption lines or high \ion{C}{4} blueshift) could show a softer optical/UV-to-X-ray SED shape than the nominal $\alpha_{\rm OX}-L_{\rm 2500\AA}$ relation. 
    \item Quasars at $z>6.5$ {display a potential} positive correlation between $\alpha_{\rm OX}$ and \ion{C}{4} blueshift, indicating a soft optical/UV-to-X-ray SED is critical for launching fast nuclear winds. We find no {significant} redshift evolution in the $\alpha_{\rm OX}$ and \ion{C}{4} blueshift relation up to $z>6.5$, suggesting {that} the mechanism for launching disk winds does not change with redshift. 
    \item We analyze the X-ray spectrum of 11 quasars at $z>6.5$ detected by Chandra through spectrum stacking and joint analysis of individual X-ray spectrum. 
    By assuming a redshifted power-law plus a galactic absorption model, we find the best-fit photon index $\Gamma$ is {$2.41\pm0.27$ ($2.67\pm0.34$)} from the stacked spectrum (joint analysis of individual X-ray spectrum). We argue the high $\Gamma$ of luminous $z>6.5$ quasars is mainly driven by their high accretion rates. 
    \item We study the correlations between {the quasar} luminosity (using $L_{\rm X}$ or $L_{\rm bol}$) and {the} star formation luminosity of the host galaxy, using $L_{\rm IR}$. We find no correlations between either $L_{\rm X}$ and $L_{\rm IR}$, nor $L_{\rm bol}$ and $L_{\rm IR}$. 
    This suggests no observed correlation between {the} quasar luminosity and {the} star formation luminosity for the most luminous quasars at $z>6.5$. 
\end{itemize}

\section*{Acknowledgments}
We thank the anonymous reviewer for their constructive comments that have improved this manuscript. 
We thank Jian Huang and Richard Green for helpful discussions.
XJ was supported by the NSF through award SOSPA9-004 from the NRAO. FW acknowledges support from NSF award AST-2513040. 
J.T.L. acknowledge the financial support from the National Science Foundation of China (NSFC) through the grants 12321003 and 12273111, and also the science research grants from the China Manned Space Program with grant no. CMS-CSST-2025-A04 and CMS-CSST-2025-A10. 
HZ was supported by NASA Astrophysics Theory Program (ATP) grant, 23-ATP23-0095.
W.L. was supported in part by NASA through STScI grant JWST-SURVEY-3428.
W.L.T acknowledged supports by NSF grants AST 19-08284. 
Y.W. acknowledges support from the National Key R\&D Program of China (grant no. 2023YFA1605600) and Tsinghua University Initiative Scientific Research Program (No. 20223080023).

This paper employs a list of Chandra datasets, obtained by the Chandra X-ray Observatory, contained in~\dataset[doi: 10.25574/cdc.486]{https://doi.org/10.25574/cdc.486}. This research has also made use of data obtained from the Chandra Data Archive provided by the Chandra X-ray Center (CXC).

We respectfully acknowledge the University of Arizona is on the land and territories of Indigenous peoples. Today, Arizona is home to 22 federally recognized tribes, with Tucson being home to the O’odham and the Yaqui. The University strives to build sustainable relationships with sovereign Native Nations and Indigenous communities through education offerings, partnerships, and community service.

This manuscript has benefited from grammar checking and proofreading using ChatGPT (OpenAI; https://openai.com/chatgpt).
%

\vspace{5mm}
\facilities{CXO, ALMA, Keck, Gemini, VLT, Magellan}


\software{Astropy \citep{astropy:2013, astropy:2018, astropy:2022}, BEHR \citep{Park2006ApJ}, CIAO \citep{Fruscione2006SPIE}, Numpy \citep{harris2020array}, linmix \citep{Kelly2007ApJ}, Scipy \citep{Virtanen2020SciPy}, Sherpa \citep{Siemiginowska2024ApJS}}




\bibliography{sample631}{}
\bibliographystyle{aasjournal}



\end{document}